\documentclass[twocolumn]{aastex6}
\usepackage{graphicx}
\usepackage{amssymb,amsfonts,amsmath,amstext,amsgen,amsopn,amsxtra,indentfirst,times}

\begin{document}

\title{The Peculiar Atmospheric Chemistry of KELT-9b}

\author{Daniel Kitzmann\altaffilmark{1}}
\author{Kevin Heng\altaffilmark{1}}
\author{Paul B. Rimmer\altaffilmark{2,3}}
\author{H. Jens Hoeijmakers\altaffilmark{1,4}}
\author{Shang-Min Tsai\altaffilmark{1}}
\author{Matej Malik\altaffilmark{1}}
\author{Monika Lendl\altaffilmark{5}}
\author{Russell Deitrick\altaffilmark{1}}
\author{Brice-Olivier Demory\altaffilmark{1}}
\altaffiltext{1}{University of Bern, Center for Space and Habitability, Gesellschaftsstrasse 6, CH-3012, Bern, Switzerland.  Emails: daniel.kitzmann@csh.unibe.ch, kevin.heng@csh.unibe.ch}
\altaffiltext{2}{University of Cambridge, Cavendish Astrophysics, JJ Thomson Ave, Cambridge CB3 0HE}
\altaffiltext{3}{University of Cambridge, MRC Laboratory of Molecular Biology, Francis Crick Ave, Cambridge CB2 OQH}
\altaffiltext{4}{Geneva Observatory, 51 chemin des Maillettes, 1290 Versoix, Switzerland}
\altaffiltext{5}{Space Research Institute, Austrian Academy of Sciences, Schmiedlstrasse 6, 8042 Graz, Austria}

\begin{abstract}
The atmospheric temperatures of the ultra-hot Jupiter KELT-9b straddle the transition between gas giants and stars, and therefore between two traditionally distinct regimes of atmospheric chemistry.  Previous theoretical studies assume the atmosphere of KELT-9b to be in chemical equilibrium.  Despite the high ultraviolet flux from KELT-9, we show using photochemical kinetics calculations that the observable atmosphere of KELT-9b is predicted to be close to chemical equilibrium, which greatly simplifies any theoretical interpretation of its spectra.  It also makes the atmosphere of KELT-9b, which is expected to be cloudfree, a tightly constrained chemical system that lends itself to a clean set of theoretical predictions.  Due to the lower pressures probed in transmission (compared to emission) spectroscopy, we predict the abundance of water to vary by several orders of magnitude across the atmospheric limb depending on temperature, which makes water a sensitive thermometer.  Carbon monoxide is predicted to be the dominant molecule under a wide range of scenarios, rendering it a robust diagnostic of the metallicity when analyzed in tandem with water.  All of the other usual suspects (acetylene, ammonia, carbon dioxide, hydrogen cyanide, methane) are predicted to be subdominant at solar metallicity, while atomic oxygen, iron and magnesium are predicted to have relative abundances as high as 1 part in 10,000.  Neutral atomic iron is predicted to be seen through a forest of optical and near-infrared lines, which makes KELT-9b suitable for high-resolution ground-based spectroscopy with HARPS-N or CARMENES.  We summarize future observational prospects of characterizing the atmosphere of KELT-9b.
\end{abstract}


\section{Introduction}
\label{sect:intro}

The highly-inflated gas-giant exoplanet KELT-9b is a standout/oddball: it is the first hot Jupiter to be found orbiting a A/B star, has an equilibrium temperature in excess of 4000 K, and a $z^\prime$-band brightness temperature of about 4600 K \citep{gaudi17}.  Based on existing empirical trends \citep{heng16,stevenson16}, we expect its atmosphere to be cloudfree.  \cite{bc18} have previously suggested that the transition between the dayside and nightside of KELT-9b may coincide with a transition between the dominance of atomic and molecular hydrogen, respectively.  This implies that the latent heat associated with transforming hydrogen from its atomic to molecular forms (and vice versa) may alter the efficiency of day-night heat redistribution.  (See also \citealt{kt18}.)  In the current Letter, we study the atomic and molecular chemistry of KELT-9b, which we believe to be peculiar as KELT-9b exhibits chemical traits of both a gas giant and a star within the same atmosphere.

Despite being irradiated in the ultraviolet about 700 times stronger than for WASP-33b \citep{gaudi17}, our expectation is that the abnormally high temperatures allow the observable atmosphere of KELT-9b to remain close to chemical equilibrium (Section \ref{sect:ce}).  An atmosphere in chemical equilibrium is considerably easier to interpret, because the atomic and molecular abundances are determined only by the temperature, pressure and elemental abundances (e.g., Chapter 7 of \citealt{heng17}).  Chemistry becomes a local, rather than global, problem.  From a qualitative standpoint, we may begin to understand the atmospheric chemistry of KELT-9b from the following net reactions (e.g., \citealt{moses11,ht16}),
\begin{equation}
\begin{split}
2\mbox{H} + \mbox{M} &\leftrightarrows  \mbox{H}_2 + \mbox{M}, \\
\mbox{CH}_4 + \mbox{H}_2\mbox{O} &\leftrightarrows \mbox{CO} + 3 \mbox{H}_2, \\
\end{split}
\label{eq:reactions}
\end{equation}
where M is a third body.  Lower and higher pressures favor atomic and molecular hydrogen, respectively.  This implies that the transition from H$_2$ to H should occur at lower temperatures in transmission versus emission spectra, since the pressure probed in transmission is a factor $\sqrt{H/R} \sim 10$ lower \citep{fortney05,hk17}, with $H$ being the pressure scale height and $R$ being the transit radius.  

Since carbon monoxide (CO) needs molecular hydrogen to make water (H$_2$O) and methane (CH$_4$), this conversion may be inhibited depending on the unknown temperature of the transit chord of KELT-9b.  If the temperature probed in transmission is close to the equilibrium temperature of about 4000 K, then we expect atomic hydrogen to reign and for water to be scarce.  It is abundant if the temperature is somewhat lower.  One of our goals is to establish quantitatively how sensitive this dependence on temperature is.  In this way, water may be used as a thermometer for the atmospheric limb temperature of KELT-9b.  Even if water is undetected, an upper limit set on its relative abundance is a lower limit on temperature (Section \ref{sect:diagnostics}).  Carbon monoxide is the dominant molecule and we expect its abundance to serve as a direct diagnostic for the elemental abundances of carbon and oxygen.

At the same time our study was released, several other studies explored the atmospheric chemistry of ultra-hot Jupiters as well.  \cite{par18} combined a three-dimensional general circulation model (GCM) with the \texttt{NASA CEA} code, which assumes chemical equilibrium, and focused on WASP-121b.  \cite{loth18} used one-dimensional \texttt{PHOENIX} models coupled to the \texttt{Astrophysical Chemical Equilibrium Solver} to study KELT-9b.  \cite{man18} confronted Hubble Space Telescope (HST) Wide Field Camera 3 (WFC3) of HAT-P-7b with both three-dimensional GCMs and one-dimensional radiative transfer models that assume chemical equilibrium.  \cite{kr18} also confronted these chemical-equilibrium models with HST-WFC3 data, but for WASP-103b.  \cite{arc18} used one-dimensional chemical-equilibrium models to confront HST-WFC3 data of WASP-18b.  The key point is that \textit{none} of these studies actually examine if chemical equilibrium is a reasonable assumption, but instead \textit{assume} it.  In the current study, we use the \texttt{ARGO} photochemical kinetics code \citep{rh16}, together with a validated stellar spectral energy distribution of KELT-9, to study this issue.

From the study of stellar atmospheres, we expect the dayside atmosphere of KELT-9b to resemble that of a K star with its spectrum being dominated by atoms \citep{gaudi17}.  What is less obvious is that we expect to see a forest of optical and near-infrared lines associated with both neutral and singly-ionized iron (e.g., \citealt{ns88,s92,ch06,aj07,peterson17}).  In Section \ref{sect:obs}, we show simulated ground-based, high-resolution spectra associated with them.  The overarching motivation of this study is to theoretically establish KELT-9b as a unique laboratory for atmospheric chemistry and thereby position it as a target for future observations across all wavelengths.

\section{Is the atmosphere of KELT-9b in chemical equilibrium?}
\label{sect:ce}

\begin{figure}
\begin{center}
\includegraphics[width=0.9\columnwidth]{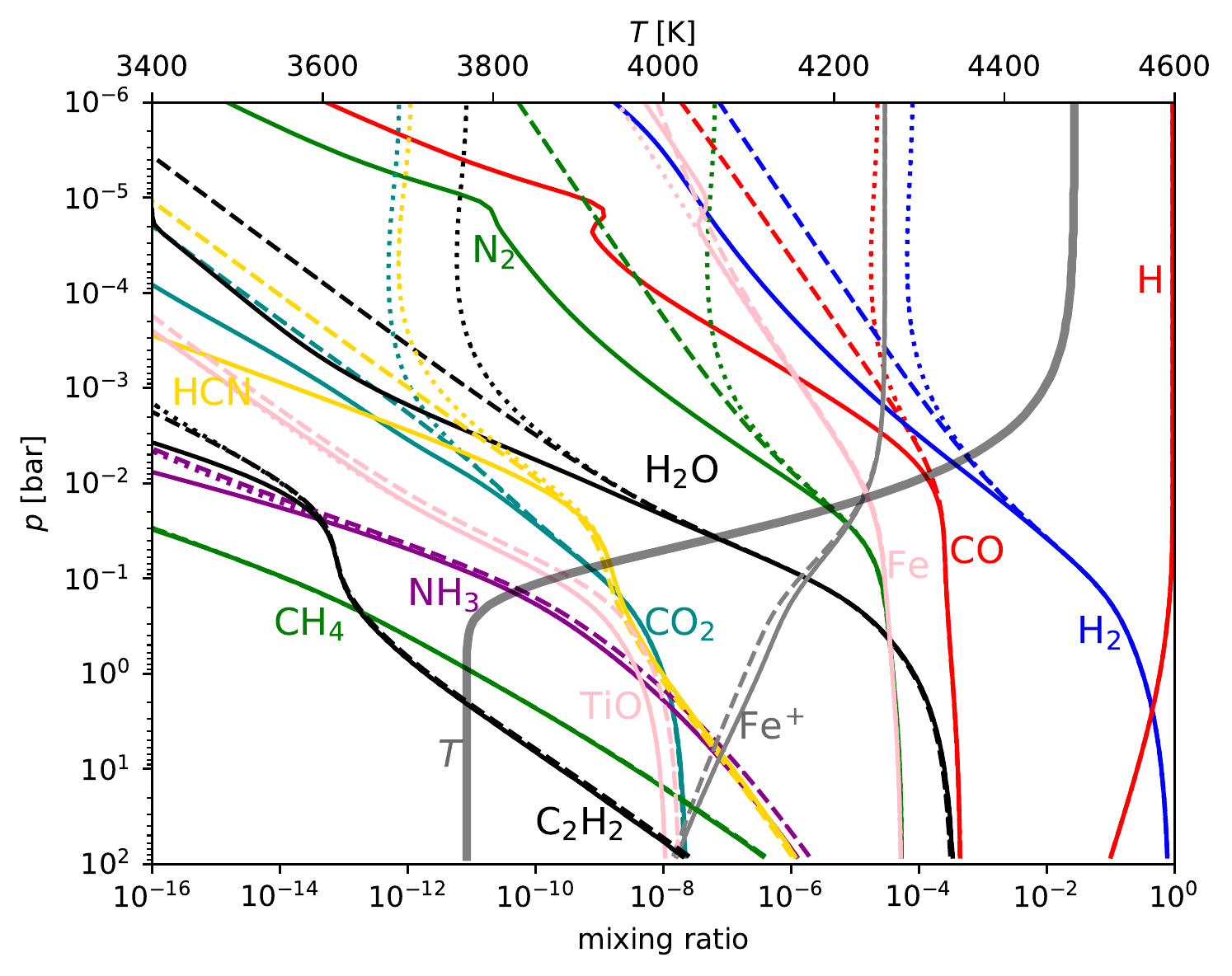}
\includegraphics[width=0.9\columnwidth]{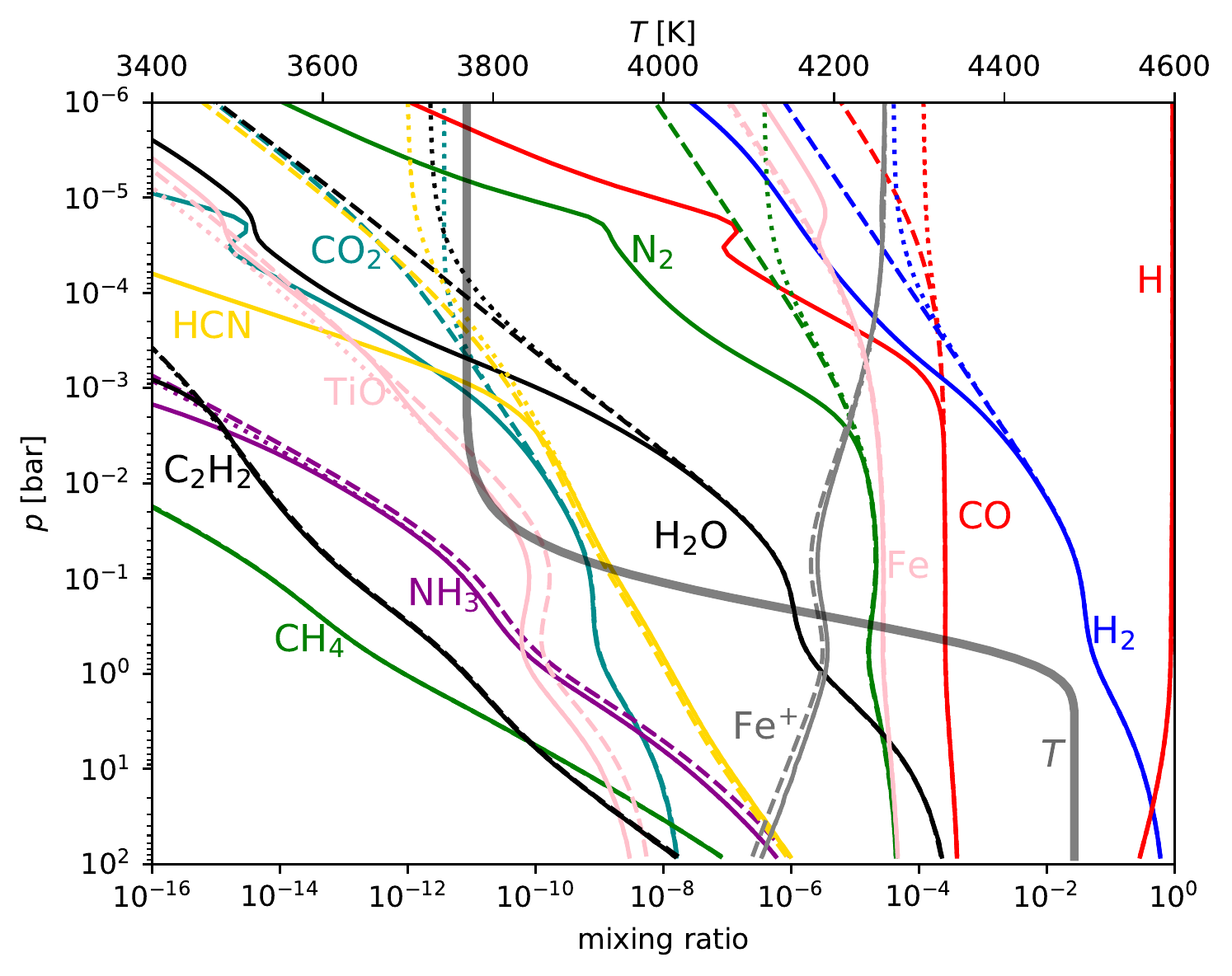}
\includegraphics[width=0.9\columnwidth]{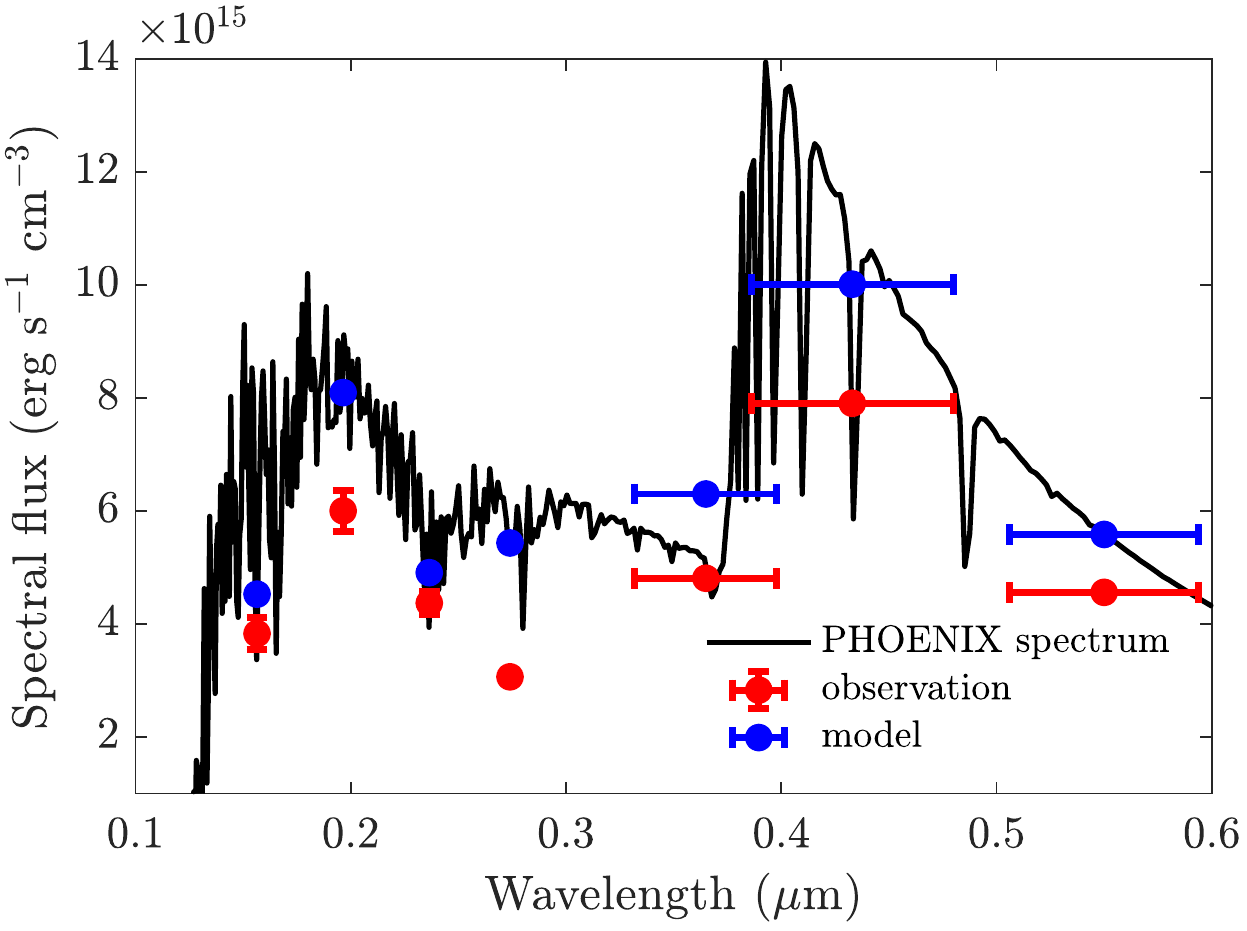}
\end{center}
\vspace{-0.1in}
\caption{Chemical kinetics calculations of model atmospheres of KELT-9b with the nominal temperature-pressure profiles overlaid.  The dashed curves assume chemical equilibrium, while the solid curves include both vertical mixing and photochemistry.  The dotted curves include vertical mixing only.  The top and middle panels assume temperature-pressure profiles with and without temperature inversions, respectively.  A vertical mixing strength of $K_{\rm zz} = 10^{10}$ cm$^2$ s$^{-1}$ is assumed.  Bottom panel: Comparing the measured ultraviolet fluxes from the stellar surface of KELT-9/HD 195689 versus those from the \texttt{PHOENIX} stellar model.  The model values are binned in the same way as the measured ones for comparison, but the entire \texttt{PHOENIX} curve is used in the photochemistry calculations.}
\label{fig:ce}
\end{figure}

\begin{figure}
\begin{center}
\includegraphics[width=\columnwidth]{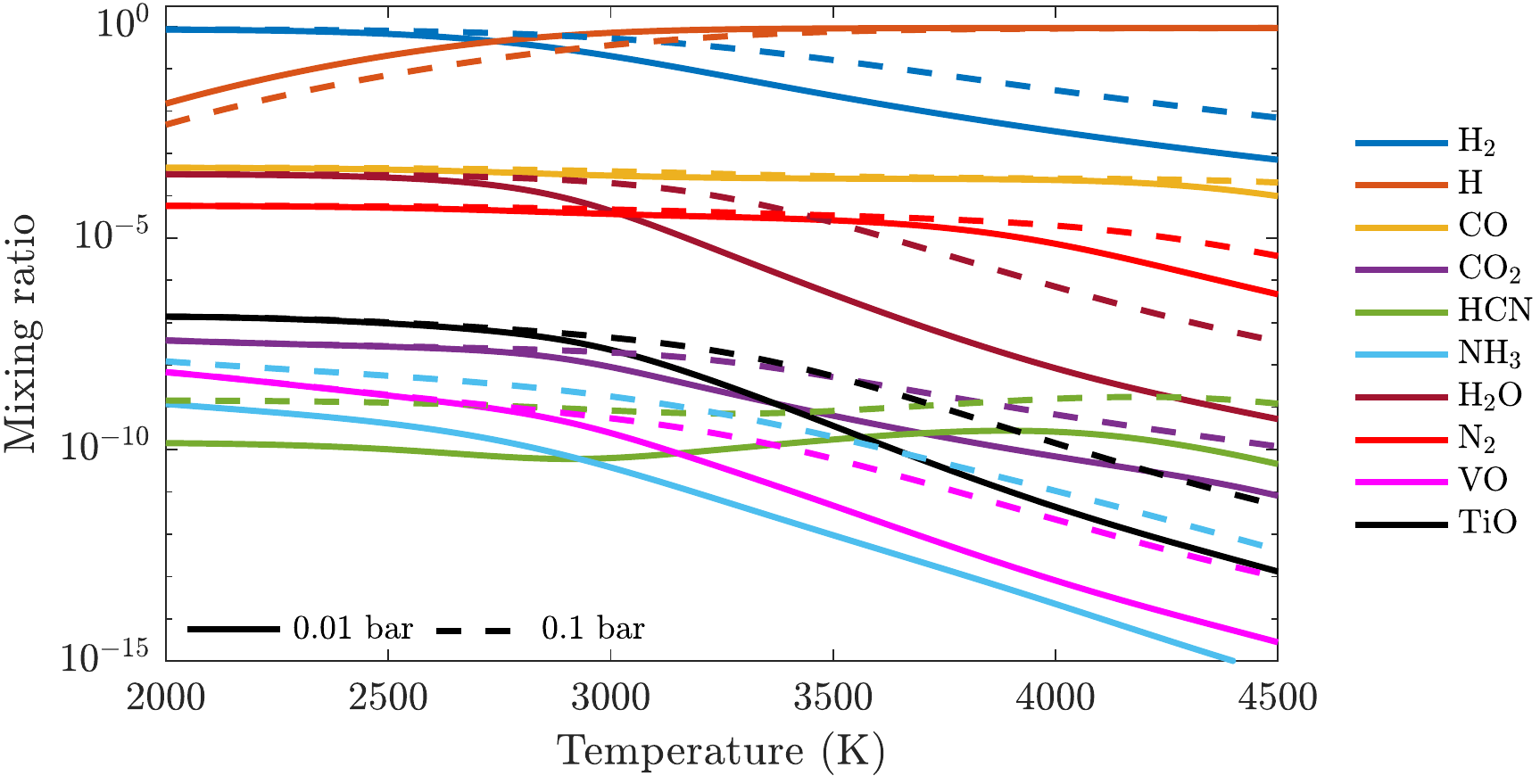}
\includegraphics[width=\columnwidth]{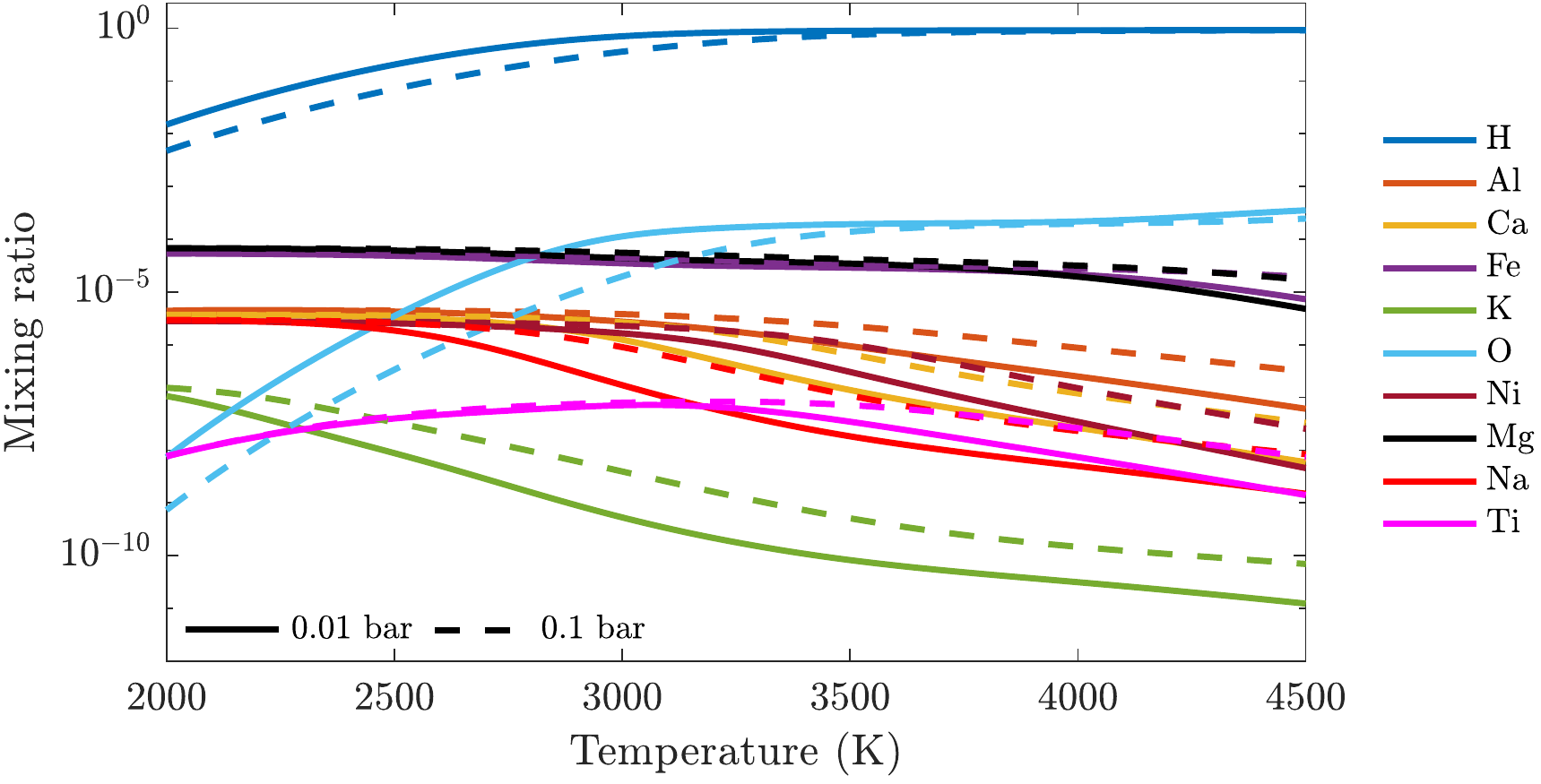}
\includegraphics[width=\columnwidth]{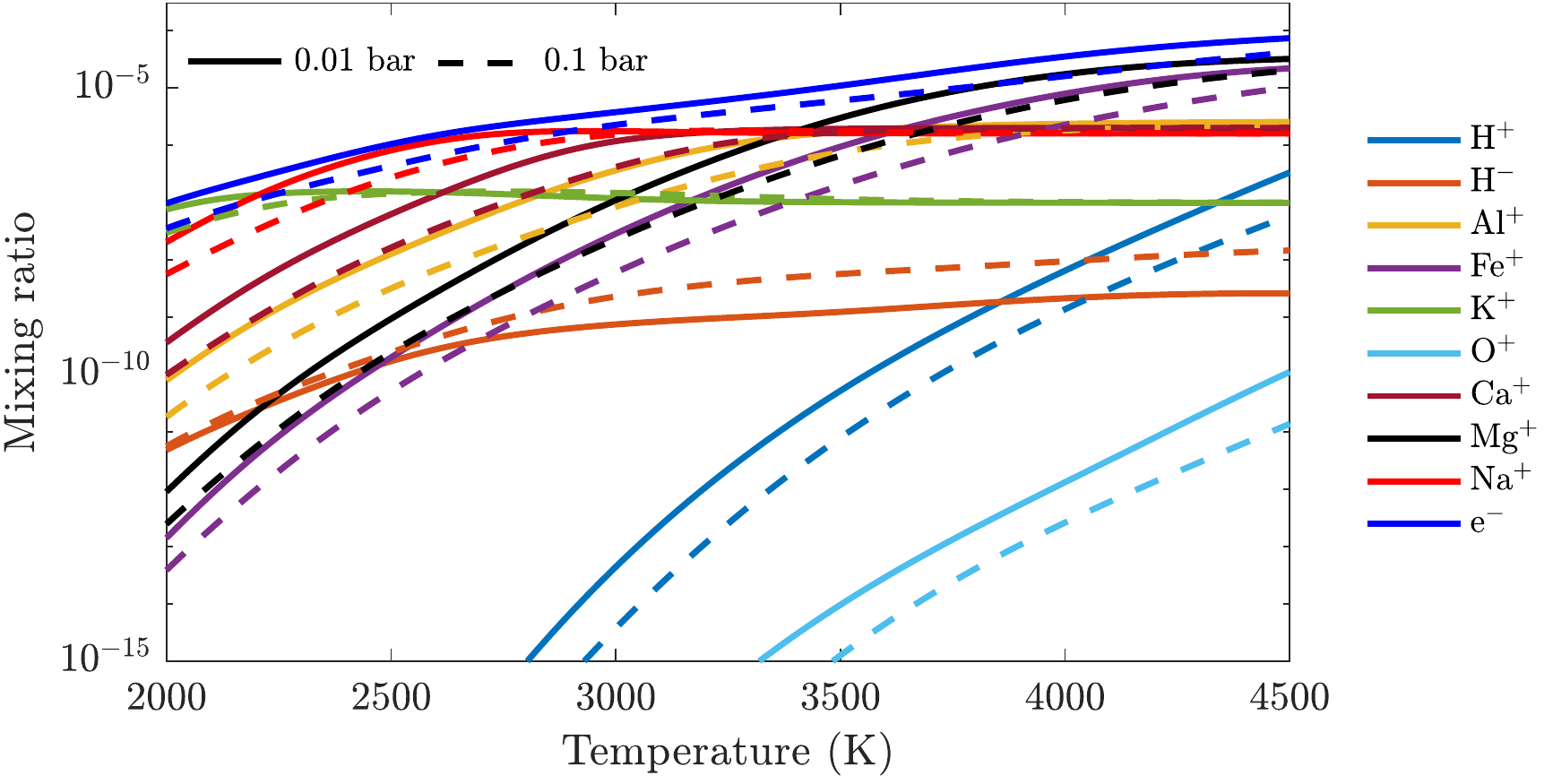}
\end{center}
\vspace{-0.1in}
\caption{Equilibrium chemistry calculations assuming solar elemental abundances.  The solid and dashed curves correspond to pressures of 10 mbar and 0.1 bar, respectively, which are meant to represent the abundances probed in transmission versus emission.  Shown are the mixing ratios of molecules (top panel), atoms (middle panel) and ions (bottom panel).}
\label{fig:chem_temp}
\end{figure}

Based on our earlier study of WASP-12b \citep{oreshenko17,tsai17}, we had some expectation that the atmosphere of KELT-9b would be close to chemical equilibrium because of the higher temperatures involved.  To check our expectation, we use the \texttt{ARGO} photochemical kinetics code \citep{rh16} to compute model atmospheres of KELT-9b.  A zenith angle of $48^\circ$ is assumed, which corresponds to its cosine being $2/3$.  Unless otherwise stated, we assume the solar metallicity values of \citet{asp09}: $\mbox{C/H}=2.69 \times 10^{-4}$, $\mbox{O/H}=4.90 \times 10^{-4}$ and $\mbox{N/H}=0.68 \times 10^{-4}$ such that $\mbox{C/O}=0.55$ and $\mbox{N/O}=0.14$.  

We assume two temperature-pressure profiles using the analytical formula of \cite{guillot10}, which only requires three parameters: the equilibrium temperature (a known quantity: 4050 K), a mean infrared opacity and a mean optical opacity.  (See also Chapter 4 of \citealt{heng17}.)  Since the hydrogen anion (H$^-$) is expected to be the dominant source of spectral continuum \citep{arc18}, we choose a mean infrared opacity of 0.01 cm$^2$ g$^{-1}$ such that the pressure probed by the transit chord is $\sim 10$ mbar, consistent with a first-principles calculation using the H$^-$ cross section for a hydrogen-dominated atmosphere \citep{john88}.  We then choose the mean optical opacity to be half and four times of the mean infrared opacity, such that profiles without and with a temperature inversion are generated, respectively.  One of these profiles is motivated by the expectation that very hot Jupiters are expected to have temperature inversions (e.g., \citealt{arc18}). 

While the use of fixed temperature-pressure profiles implies that our photochemical kinetics and radiative transfer calculations are not self-consistent, they are justified, as a first approach, for three reasons.  First, our primary aim is to examine the role of photochemistry.  A predominance of the currently available photochemical cross sections do not depend on temperature, even if in principle they should \citep{venot13}.  Second, this non-self-consistent approach is prevalent in the published literature (e.g., \citealt{line11,moses11,moses13,vis12,zm14,venot15,tsai17}).  Third, self-consistency between chemical kinetics and radiative transfer is expected to drive the system \textit{closer} to chemical equilibrium \citep{drummond16}, implying that our approach is conservative.

We paid close attention to the ultraviolet fluxes of KELT-9 reported by \cite{gaudi17}.  We use a \texttt{PHOENIX} stellar model \citep{Husser} interpolated to the stellar parameters of KELT-9 reported in \cite{gaudi17}.  We then binned the model ultraviolet fluxes and compared them to the measured values (Figure \ref{fig:ce}).  We see that our \texttt{PHOENIX} model ultraviolet fluxes are $\sim 10\%$ higher than the KELT-9 measurements, implying that the ultraviolet spectrum of KELT-9 we are using for our photochemical kinetics calculations is conservative.  Being an A/B star, KELT-9 is expected to be fully radiative and not possess a magnetic dynamo that drives X-ray emission \citep{ss07}.  As such, we do not need to concern ourselves with X-ray emission from KELT-9 impinging upon the atmosphere of KELT-9b and influencing the photochemistry.

\cite{cs06} used a three-dimensional GCM coupled to a chemical relaxation method (dealing only with methane-to-carbon-monoxide interconversion), which assumes that chemical abundances relax towards their chemical-equilibrium values on a prescribed timescale, and found that vertical mixing is more important than horizontal mixing for disequilibrium chemistry.  A Newtonian relaxation scheme was used in lieu of radiative transfer.  \cite{agundez14} found horizontal mixing to be more important, but their conclusion is based on a simplified two-dimensional dynamical model that assumes the zonal wind to follow solid-cylinder rotation, although they do perform full chemical kinetics calculations.  \cite{drummond18} coupled a three-dimensional GCM to the chemical relaxation method and two-stream radiative transfer and reach a more subdued conclusion than \cite{agundez14} regarding the importance of horizontal mixing.  The studies of \cite{cs06} and \cite{drummond18} predate that of \cite{tsai18}, who tested the accuracy of the chemical relaxation method for the first time.  \cite{tsai18} showed that, contrary to the implementation of \cite{cs06} and \cite{drummond18}, a single chemical timescale corresponding to a single rate-limiting reaction cannot be assumed, and that order-of-magnitude differences in the chemical timescale exist between the approximation of \cite{cs06} and more accurate calculations using full chemical kinetics.  Given this evolving literature, our use of one-dimensional, plane-parallel models of chemical kinetics with photochemistry and vertical mixing only, to study whether the atmosphere of KELT-9b is in chemical equilibrium, is not unreasonable as a first approach.

In steady state, the conservation of mass states that the horizontal and vertical velocities are related by $v_h / R \sim v_z / H$.  The same argument implies that the dynamical timescales associated with vertical and horizontal mixing are comparable at the order-of-magnitude level.  Since $v_h \sim c_s$ (with $c_s$ being the sound speed), we expect that $v_z \sim v_h H / R \sim 0.01 c_s$.  This implies that the eddy diffusion coefficient associated with vertical mixing is $K_{\rm zz} \sim 0.1 H v_z \sim 10^{10}$ cm$^2$ s$^{-1}$, where we have approximated the characteristic length scale as $0.1 H$ \citep{smith98}.

Figure \ref{fig:ce} shows the chemical abundances for three atmospheric models of KELT-9b: with photochemistry and vertical mixing, with vertical mixing only and in chemical equilibrium.  Regardless of whether the temperature-pressure profile has an inversion or not, we see that vertical mixing is relatively unimportant (i.e., the dotted and dashed curves mostly coincide) compared to photochemistry.  Photochemistry drives some of the atomic and molecular abundances away from chemical equilibrium at $\lesssim 1$ mbar, but the model atmosphere remains close to chemical equilibrium at pressures of $\gtrsim 1$ mbar, which are the pressures conceivably probed by infrared and optical spectroscopy at low spectral resolution.  When a temperature inversion is present, the higher temperatures at photospheric pressures negate the disequilibrium effects of photochemistry.  If the temperatures are above 4000 K, then the molecules start to \textit{thermally} dissociate into their constituent atoms and ions.  

Although various species are in chemical disequilibrium at low pressures or high altitudes, they reside in the optically thin part of the atmosphere, implying that the assumption of chemical equilibrium remains reasonable for regions of the atmosphere that are observed in the optical and infrared range of wavelengths---at least, at low spectral resolution.

\section{The peculiar atmospheric chemistry of KELT-9b}
\label{sect:diagnostics}

\begin{figure}
\begin{center}
\includegraphics[width=\columnwidth]{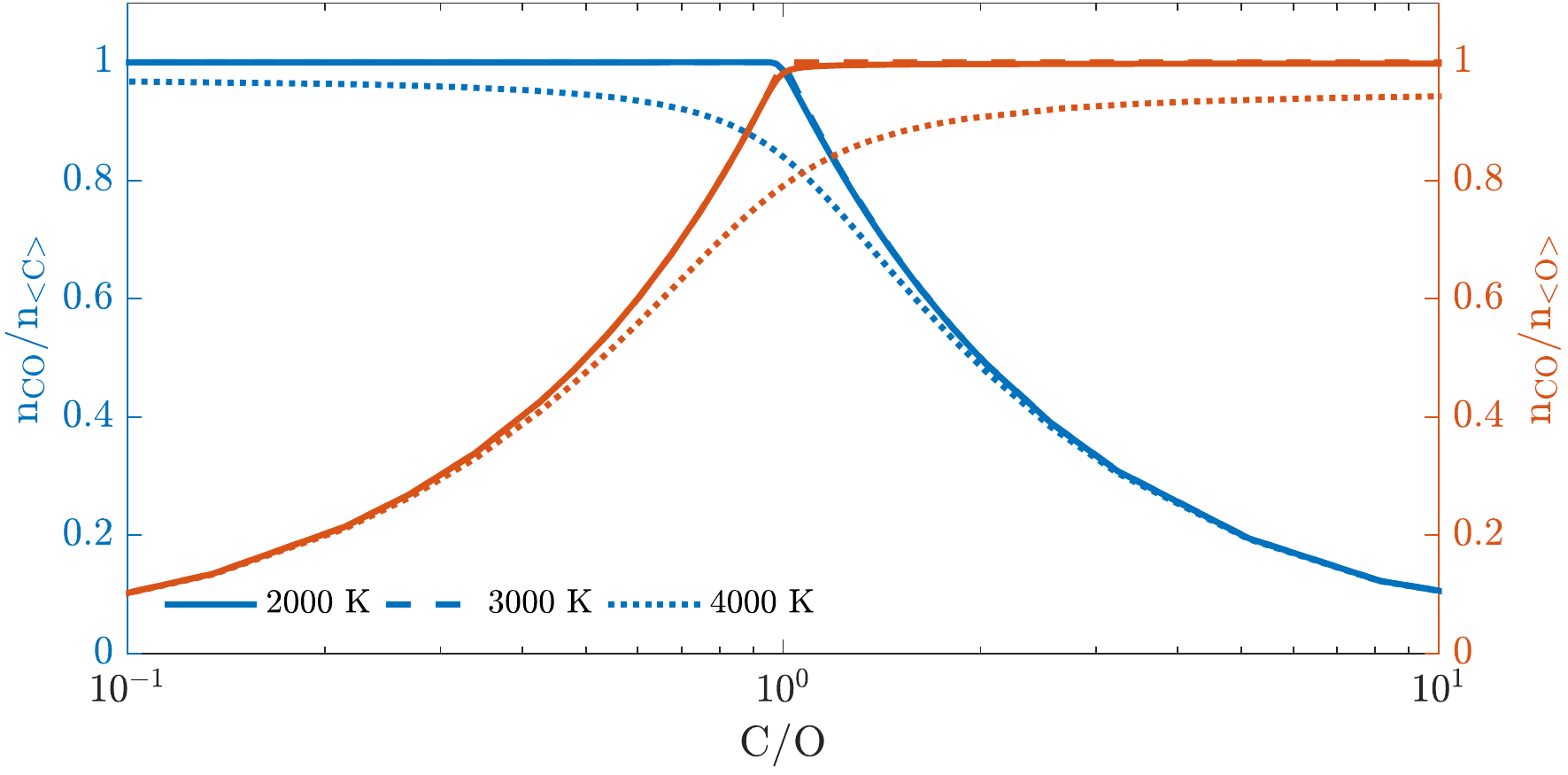}
\includegraphics[width=\columnwidth]{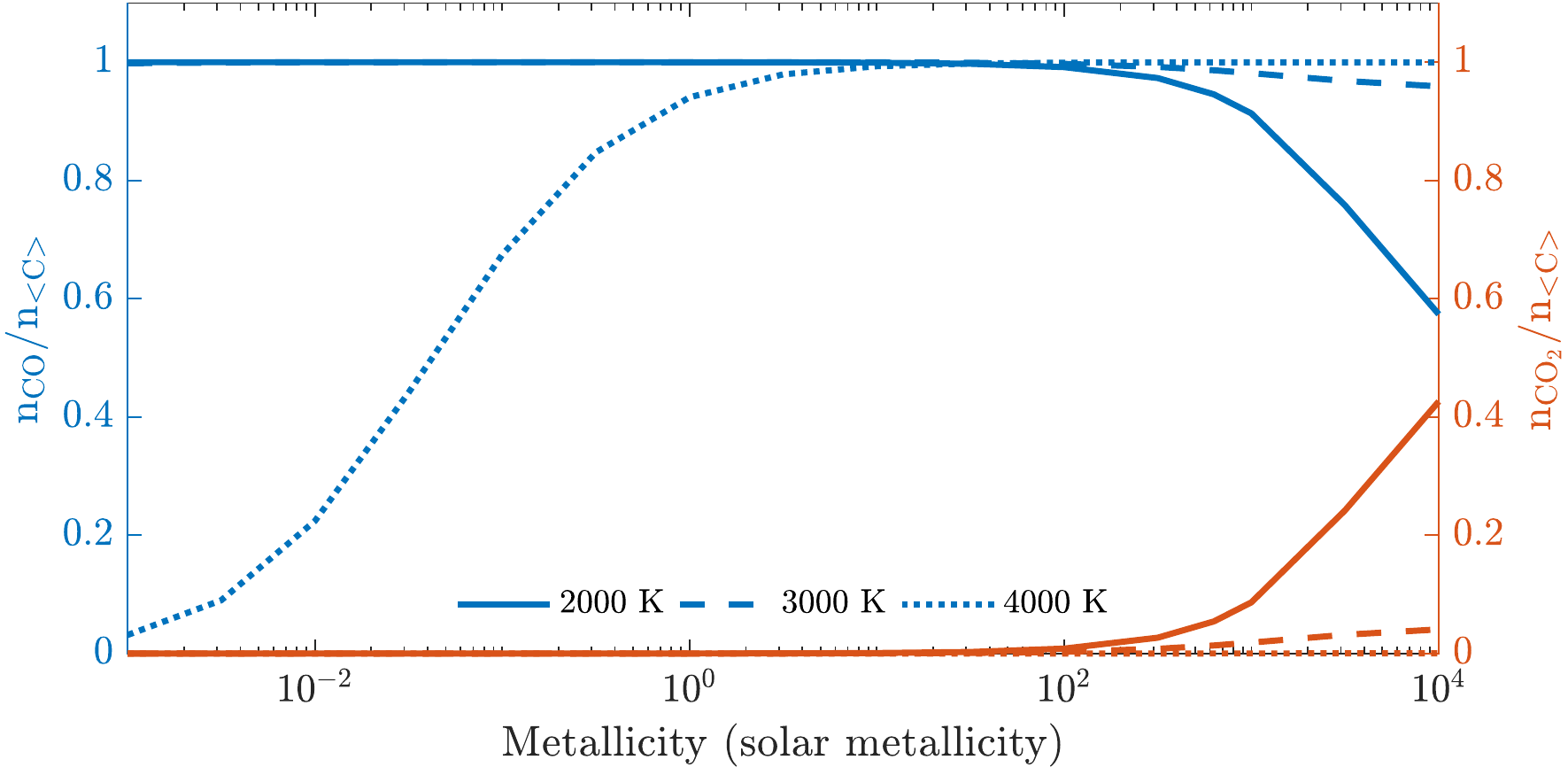}
\end{center}
\vspace{-0.1in}
\caption{Investigating CO as a metallicity diagnostic.  Top panel: number density of CO normalized by the number density of either carbon or oxygen atoms as a function of C/O.  The left plot fixes O/H to its solar value and varies C/H, while the right plot fixes C/H to its solar value and varies O/H.  Bottom panel: number density of CO and CO$_2$, normalized by the number density of carbon atoms as a function of the metallicity.  The ratio of C/H to O/H is held fixed at its solar value, but are each allowed to decrease or increase in lockstep.  A pressure of 10 mbar is assumed.}
\label{fig:co}
\end{figure}

\begin{figure*}
\begin{center}
\includegraphics[width=\columnwidth]{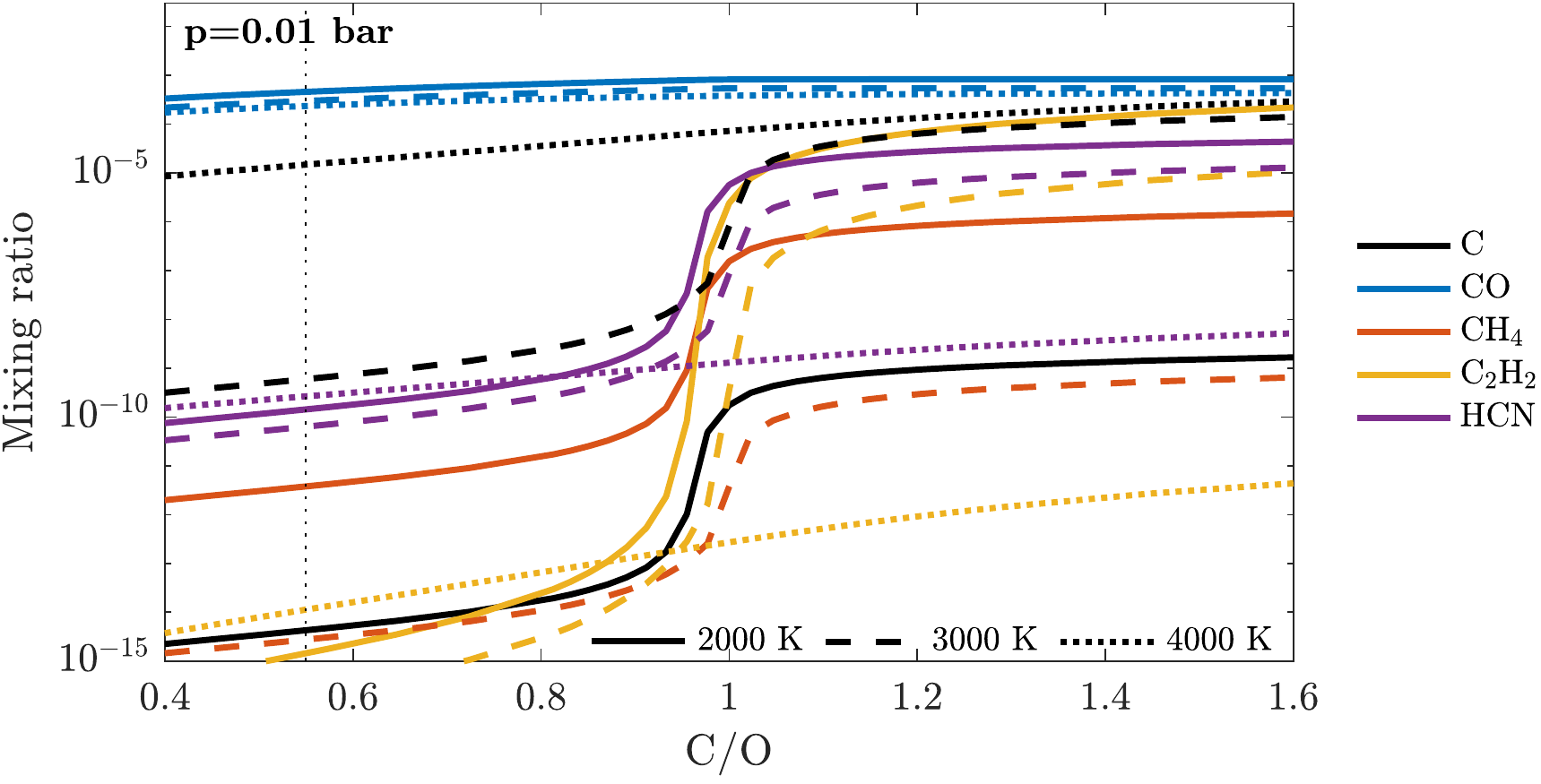}
\includegraphics[width=\columnwidth]{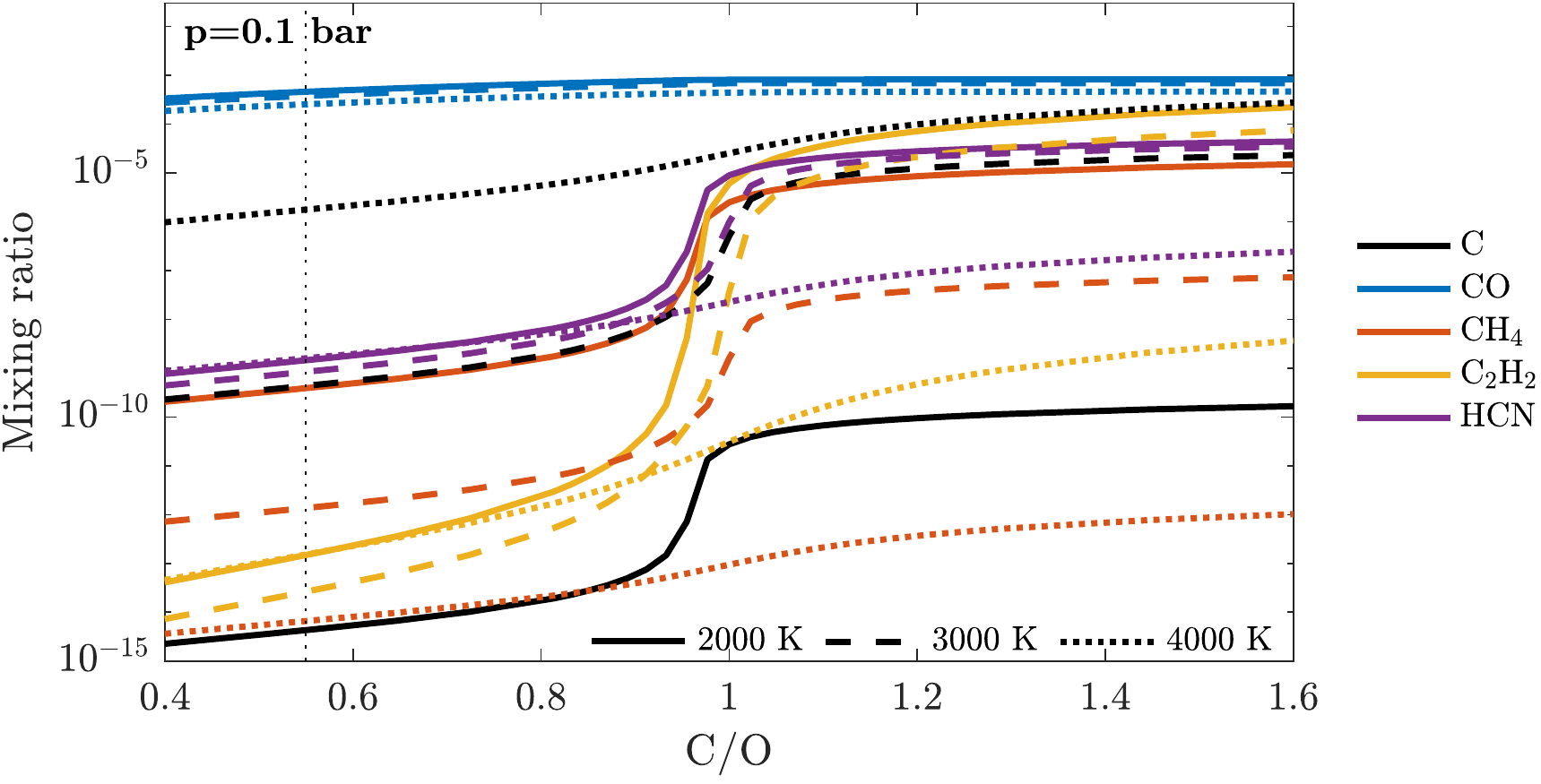}
\includegraphics[width=\columnwidth]{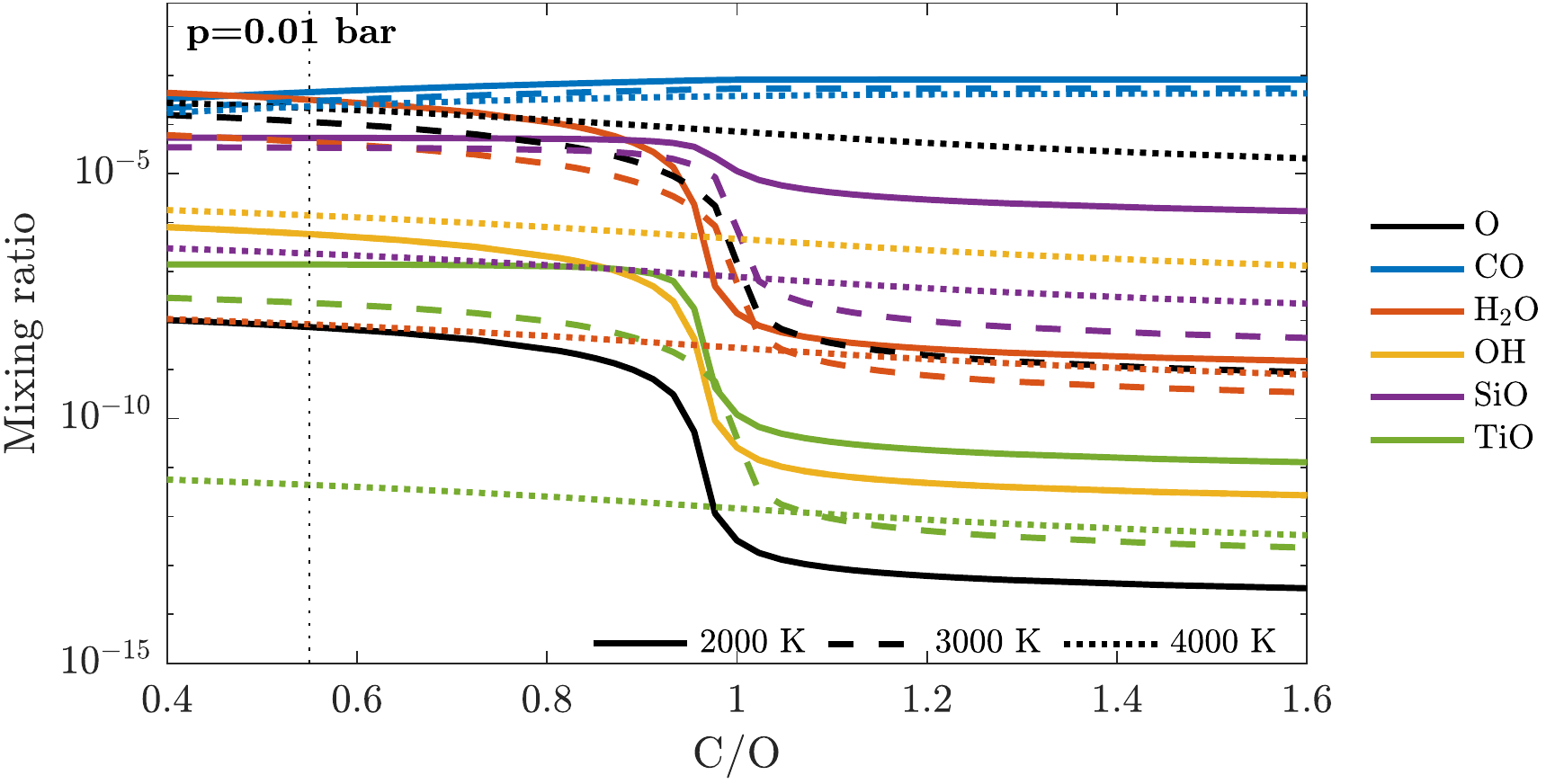}
\includegraphics[width=\columnwidth]{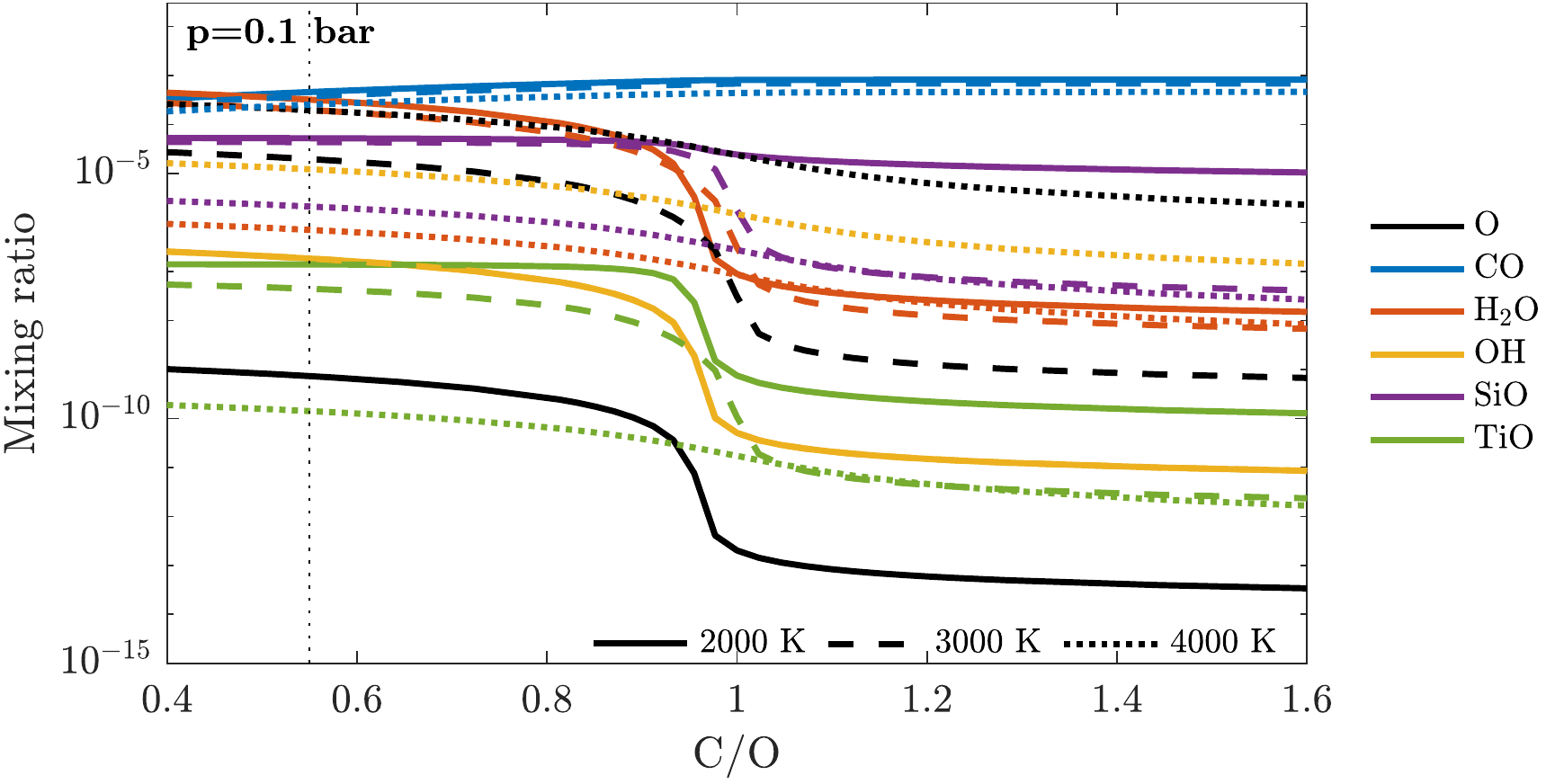}
\end{center}
\vspace{-0.1in}
\caption{Equilibrium chemistry calculations assuming a solar O/H value and allowing C/H (and hence C/O) to vary. The mixing ratios of important carbon and oxygen-bearing species are shown in the top and bottom rows, respectively.  The solid, dashed and dotted curves correspond to temperatures of 2000 K, 3000 K and 4000 K, respectively, which are meant to show the possible range of atmospheric temperatures for KELT-9b.  The left and right columns assume pressures of 10 mbar and 0.1 bar, respectively.  The vertical, dotted lines indicate the solar C/O value in each panel.  We have not shown the abundances of the iron species, because they are insensitive to C/O.}
\label{fig:chem_co}
\end{figure*}

To further explore the atmospheric chemistry (and for reasons of computational efficiency), we switch to using \texttt{FastChem}, which is an open-source chemical-equilibrium code\footnote{\texttt{https://github.com/exoclime/fastchem}} that treats gas-phase chemistry associated with the elements Al, Ar, Ca, Cl, Co, Cr, Cu, F, Fe, H, He, K, Mg, Mn, Na, Ne, P, S, Si, Ti, V and Zn, including ions \citep{stock18}.

Figures \ref{fig:chem_temp}, \ref{fig:co} and \ref{fig:chem_co} show the volume mixing ratios (relative abundances by number) of molecules, atoms and ions.  For simplicity and ease of interpretation, we approximate the pressures probed in transmission and emission to be 10 mbar and 0.1 bar, respectively.  We then explore trends associated with temperature and carbon-to-oxygen ratio (C/O).  Some general trends are worth pointing out.  First, molecules that are usually prominent in hot Jupiters are subdominant, including acetylene (C$_2$H$_2$), ammonia (NH$_3$), carbon dioxide (CO$_2$), hydrogen cyanide (HCN) and methane (CH$_4$), which have mixing ratios $\ll 10^{-6}$.  We have not explicitly shown the mixing ratios for methane and acetylene in Figure \ref{fig:chem_temp}, because their abundances mostly fall below the lower boundary of the plot ($10^{-15}$) at these high temperatures and solar metallicity.   Second, atomic oxygen (O), magnesium (Mg) and iron (Fe) may have mixing ratios as high as $\sim 10^{-4}$, which is bested only by CO and sometimes water (H$_2$O).  Even nickel (Ni) may attain mixing ratios $\sim 10^{-6}$.  Intriguingly, titanium (Ti) may be as abundant as $\sim 10^{-7}$.

\subsection{Carbon monoxide as a metallicity diagnostic}

In Figures \ref{fig:chem_temp} and \ref{fig:chem_co}, we see that the abundance of CO remains remarkably constant, across a broad range of temperatures (2000--4500 K), C/O values and for pressures of 0.01--0.1 bar.  It is not only the dominant carbon carrier by several orders of magnitude in mixing ratio (its nearest competitor is CO$_2$), but it is the dominant molecule alongside water at low temperatures and C/O values.

In Figure \ref{fig:co}, we further investigate the role of CO as a metallicity diagnostic.  At temperatures of 2000--3000 K, almost all of the available carbon or oxygen is locked up in CO and the ratio of the CO number density to the total amount of either atomic C or O is very close to unity, depending on the value of C/O.  This correspondence starts to break down at 4000 K and hotter, as CO dissociates into its constituents.  Measuring CO alone would not constrain C/H or O/H directly, because C/O is a priori unknown.  Additional molecules such as water are needed to set constraints on C/O being smaller or larger than unity and therefore to fully utilize CO as a metallicity diagnostic.  In Figure \ref{fig:co}, we also keep the \textit{ratios} of the elemental abundances fixed, but increase the elemental abundances themselves by the same factor.  For temperatures of 2000--3000 K and up to about $100 \times$ the solar metallicity, CO is a good tracer of the metallicity.  By contrast, CO$_2$ is absent at these temperatures unless the metallicity approaches 1000--10,000 times solar and the temperature is 2000 K or lower.  We consider these conditions to be unlikely for the atmosphere of KELT-9b.  We predict CO, rather than CO$_2$ (e.g., \citealt{moses13}), to play the role of the metallicity diagnostic.

\subsection{Water as a thermometer}

In emission, the abundance of water holds steady until about 3000 K or higher, beyond which it starts to dissociate (Figure \ref{fig:chem_temp}).  A less intuitive outcome is that, because of the lower pressure probed in transmission, the drop in the water abundance is much steeper across temperature.  In fact, at solar metallicity, it drops by more than 6 orders of magnitude between 2000--4500 K.  Such a sensitivity to temperature implies that water may be used as a thermometer in the transmission spectrum of KELT-9b.  Used in tandem with carbon monoxide, the elemental abundances of carbon and oxygen may be robustly determined.  The atmosphere of KELT-9b becomes a tightly constrained chemical system.

A caveat is that water may also become scarce when C/O approaches unity \citep{ks05,madhu12}, as shown in Figure \ref{fig:chem_co}.  However, when $\mbox{C/O} \gtrsim 1$, we expect acetylene, methane and hydrogen cyanide to become important.  These species are spectroscopically active in the infrared range of wavelengths and should be detectable under these conditions.  At these high temperatures, graphite formation \citep{moses13} is unimportant, which would in principle allow $\mbox{C/O}>1$ in the gas phase.

\subsection{Neutral atoms and heavy-metal content}

A prediction of having a very hot atmosphere is that the neutral species of atomic metals become prominent.  At solar metallicity, magnesium and iron have mixing ratios $\sim 10^{-4}$ in both emission and transmission from 2000--3500 K, dropping to $\sim 10^{-5}$ in transmission (with a slower drop in emission).  Above 3000 K in transmission, atomic oxygen attains mixing ratios $\sim 10^{-4}$.  Aluminum, nickel and sodium have mixing ratios as high as $\sim 10^{-6}$.  All of these properties imply that it is possible to directly constrain the heavy-metal content of the atmosphere of KELT-9b.  We will make predictions for detecting iron via ground-based, high-resolution spectroscopy in Section \ref{sect:obs}.  

\subsection{Thomson scattering by electrons versus extinction by hydrogen anions}

Another prediction is that free electrons have mixing ratios as high as $\sim 10^{-4}$ (Figure \ref{fig:chem_temp}).  For temperatures from 2000--4500 K, the mixing ratio of free electrons is comparable to or greater than that of hydrogen anions (H$^-$).  However, Thomson scattering has a somewhat small, wavelength-independent cross section of $\sigma_{\rm T} \approx 6.65 \times 10^{-25}$ cm$^2$.  The cross section associated with hydrogen anions ($\sigma_{\rm H^-}$) has a bound-free component up to 1.6419 $\mu$m, the photo-detachment threshold and a free-free component that extends into the mid-infrared range of wavelengths \citep{john88}.  Figure 1 of \cite{arc18} illustrates this graphically, where we see that  $\sigma_{\rm H^-} \sim 10^{-17}$--$10^{-16}$ cm$^2$.  We conclude that Thomson scattering is a negligible effect compared to the extinction of light by H$^-$.

\section{Opportunities for future observations}
\label{sect:obs}

\subsection{Ground-based high-resolution spectroscopy of neutral iron lines}

\begin{figure*}
\begin{center}
\includegraphics[width=2\columnwidth]{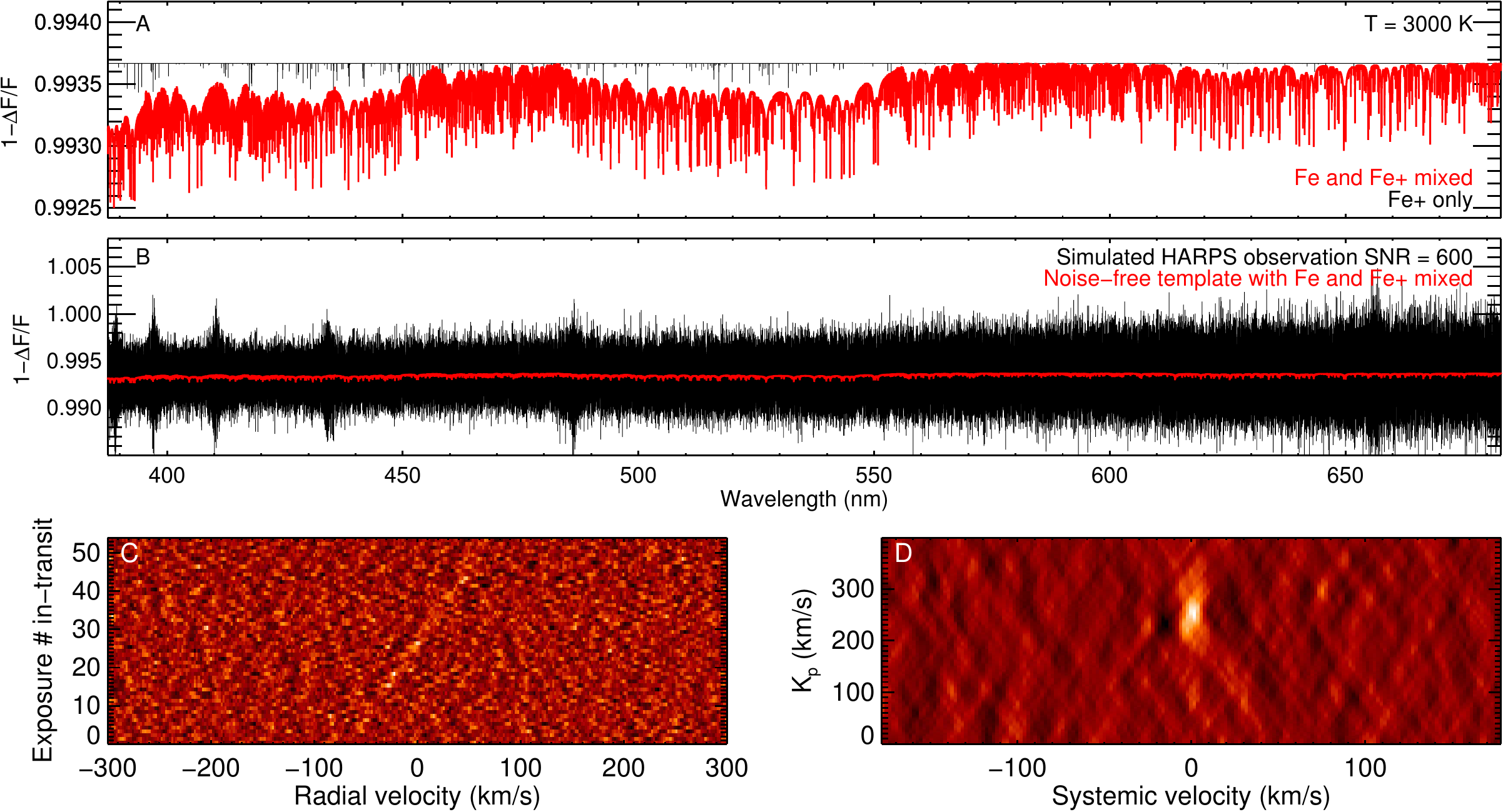}
\end{center}
\caption{Simulated HARPS-N transit observations of KELT-9b.  Panel A: theoretical transmission spectra with Fe and Fe$^+$ versus Fe$^+$ only.  Panel B: template in Panel A modulated by Gaussian noise.  The noise-free template (red curves in both Panels A and B) are drowned out by the noise, necessitating cross-correlation to co-add the individual irons lines.  Panel C: cross-correlation function (CCF) over the course of the transit.  The peak of the CCF displays a time-dependent Doppler shift due to the changing radial component of the orbital velocity of the exoplanet.  Panel D: co-addition of the CCFs in the rest frame of the exoplanet.}
\vspace{0.1in}
\label{fig:harps}
\end{figure*}

To investigate the predicted prominence of spectral lines associated with neutral and singly-ionized iron, we now simulate ground-based, resolution $\sim 100,000$ observations of them.  This technique has previously been established and consists of cross-correlating the observed system (star plus exoplanet) spectrum with a template of the atom or molecule being sought \citep{snellen10,brogi12}.  To construct this template, we use the spectroscopic line list of Fe and Fe$^+$ from the Vienna Atomic Line Database \citep{ryabchikova15}.  We compute opacities from this line list using standard methods (e.g., \citealt{grimm15,heng17}).  To generate transmission spectra, we use the analytical formula of \cite{hk17}.  For illustration, we assume a constant opacity of 0.01 cm$^2$ g$^{-1}$ to mimic a spectral continuum, a reference pressure of 0.1 bar at the white-light radius measured by \cite{gaudi17}, a transit chord (terminator) temperature of 3000 K and volume mixing ratios of of $5 \times 10^{-5}$ and $10^{-7}$ for Fe and Fe$^+$, respectively.

Figure \ref{fig:harps} shows our theoretical transmission spectra containing Fe and Fe$^+$ versus Fe$^+$ only in the visible range of wavelengths.  At the expected temperatures of KELT-9b, the energy levels of Fe are preferentially populated compared to Fe$^+$.  It is therefore unsurprising that Fe is dominant over Fe$^+$ as an opacity source.  We expect the star to be deficient in Fe lines, because calculations assuming collisional ionization equilibrium have shown that the abundance of Fe drops rapidly as the temperature exceeds 10,000 K; the abundance of Fe$^+$ becomes dominant only at these temperatures \citep{k08}.

In Figure \ref{fig:harps}, we also show a simulated transit observation of KELT-9b with the HARPS-N spectrograph located at the 3.6-meter TNG telescope at La Palma, which has a spectral resolution of about 110,000. Due to the brightness of KELT-9, HARPS-N is expected to provide a signal-to-noise ratio (SNR) of at least 100 in the stellar continuum at 500 nm after a single exposure of 200 s.  The transit duration is 3 hours, implying that 54 observations may be obtained in-transit, resulting in a combined SNR of 735 at 500 nm.  Here, our \texttt{PHOENIX} model for KELT-9 \citep{Husser} is convolved with the line-spread function to match the rotational velocity broadening ($v \sin{i}=111.4$ km s$^{-1}$), because we are interested in resolving individual spectral lines.  We also multiplied it with a \texttt{SkyCalc} model from ESO \citep{noll12,jones13} of the telluric transmission spectrum across the wavelength range accessible by HARPS-N (387--691 nm).  Our prediction is that iron can be observed at high signal-to-noise after observing a single transit with a high-resolution spectrograph on a 4-meter-class telescope, e.g., with HARPS-N or CARMENES.

\subsection{Hubble Space Telescope (HST)}

Ultraviolet transit spectroscopy using HST is now an established technique (e.g., \citealt{ehrenreich15}).  It has been used to detect both atomic hydrogen and metals (e.g., \citealt{vm04}).  Besides the semi-forbidden lines, there is a zoo of near-ultraviolet \textit{resonant} lines associated with neutral and singly-ionized metal atoms, e.g., see Tables 9.3 and 9.4 of \cite{draine}.  These include the Fe I doublet at 0.37 and 0.39 $\mu$m, a set of Ni I lines from 0.34--0.35 $\mu$m, a set of Ti I lines from 0.33--0.52 $\mu$m, the C I triplet at 0.17 $\mu$m, the O I triplet at 0.13 $\mu$m and a Mg I line at 0.29 $\mu$m.

\subsection{James Webb Space Telescope (JWST)}

Our predictions may be decisively tested by JWST via the measurement of the transmission spectrum, dayside emission spectrum and nightside emission spectrum.  Such a program will conceivably fit within a small JWST proposal ($\le 25$ hours).  Detecting CO and H$_2$O jointly will set constraints on the temperatures, C/O and metallicity.

\subsection{Characterizing Exoplanet Satellite (CHEOPS)}

\begin{figure}
\begin{center}
\includegraphics[width=\columnwidth]{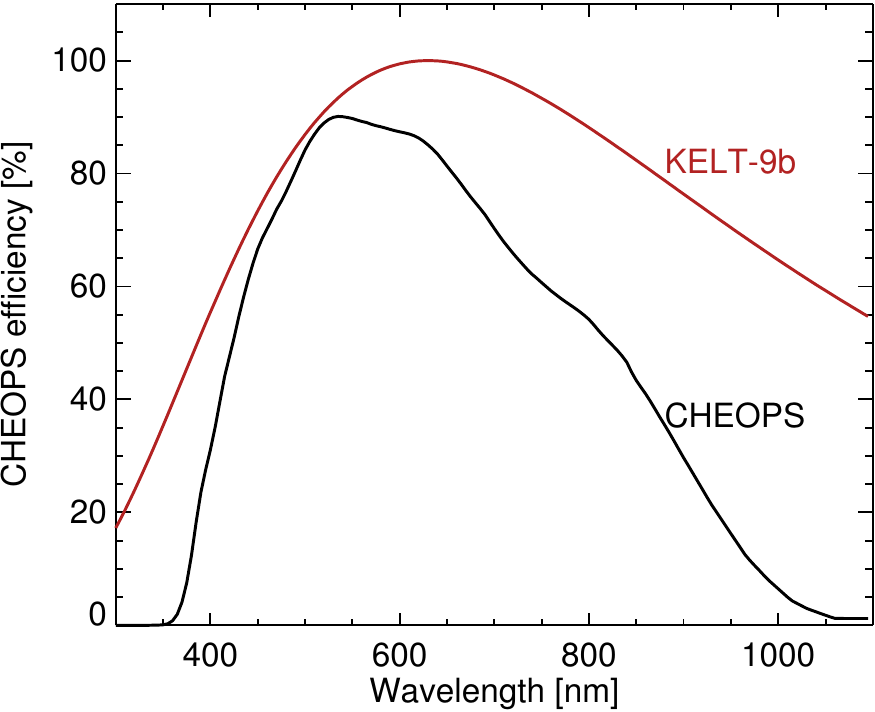}
\includegraphics[width=\columnwidth]{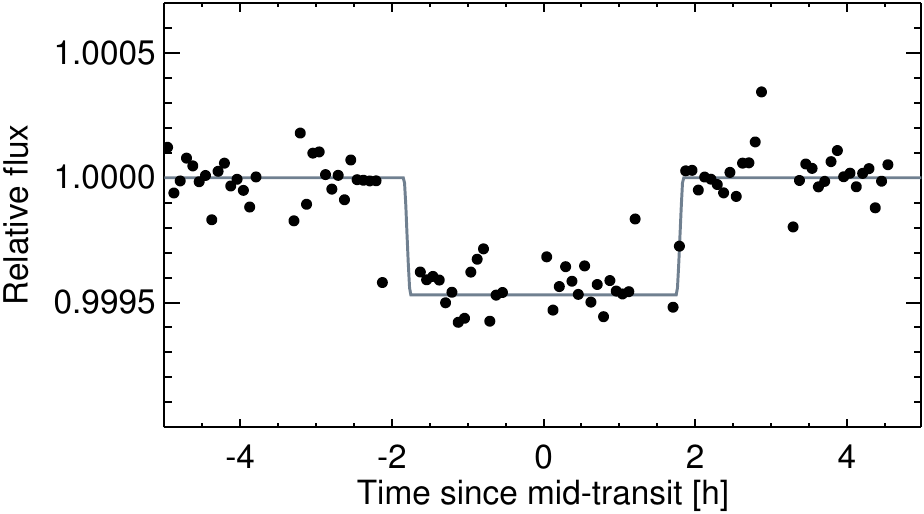}
\end{center}
\caption{Top panel: blackbody flux from KELT-9b assuming a dayside temperature of 4600 K, following the $z^\prime$-band brightness temperature of the same value.  The CHEOPS bandpass is also shown.  Bottom panel: simulated secondary eclipse of KELT-9b measured using CHEOPS.}
\vspace{0.1in}
\label{fig:cheops}
\end{figure}

The CHEOPS bandpass covers 0.33--1.1 $\mu$m \citep{broeg13}.  For hot Jupiters with equilibrium temperatures $\lesssim 1300$ K, a CHEOPS secondary eclipse translates directly into a measurement of the geometric albedo \citep{seager10}.  For somewhat hotter temperatures, the secondary eclipse measurement samples a combination 
of reflected light and thermal emission that depends on the dayside-nightside heat redistribution efficiency \citep{hd13}.  The dayside of KELT-9b is so hot that its blackbody peak wavelength is about 0.65 $\mu$m, which is in the  middle of the CHEOPS bandpass (Figure \ref{fig:cheops}).  Similar to HST, CHEOPS will reside in low-Earth orbit, meaning Earth crossings of the South Atlantic Anomaly will prevent continuous monitoring in any direction.  KELT-9 is visible to CHEOPS for $\approx 2.5$ months per year and observations can be obtained at an observing efficiency between 50--70\% (with the upper limit being the maximum obtainable), yielding about 35 occultations annually.   

The stellar effective temperature is 10,150 K, while for the planetary equilibrium temperature we assume no redistribution ($f^\prime=2/3$ in equation [3.9] of \citealt{seager10}) and a Bond albedo of 0.1 (geometric albedo of 0.067 assuming isotropic scattering).  Both the star and exoplanet are assumed to emit as blackbodies, and we integrate their fluxes in the CHEOPS bandpass, which is shown in Figure \ref{fig:cheops}.  Using equations (3.40) and (3.49) in \cite{seager10}, we compute the secondary eclipse depths due to thermal emission and reflected light, respectively.  For KELT-9b, these are 427.9 parts per million (ppm) and 45.6 ppm, which gives 473.5 ppm in total.  The thermal emission dominates reflected light by a factor of more than 9, implying that CHEOPS secondary eclipses are unambiguously sampling the thermal emission from KELT-9b.  

To compute the precision attained with a single secondary eclipse of KELT-9b, we use the CHEOPS Exposure Time Calculator (ETC).\footnote{\texttt{https://www.cosmos.esa.int/web/cheops-guest\\-observers-programme/open-time-workshop-2017}}  As input, we use the properties of the KELT-9 star \citep{gaudi17}.  The exposure time was set to 1 second.  Stellar granulation noise was excluded from the calculation as the star does not have a convective envelope.  With these inputs, we estimate a precision of about 23 ppm for a single secondary eclipse (Figure \ref{fig:cheops}).  This translates into a signal-to-noise ratio of $\approx 21$, equivalent to better than 5\% precision in the thermal emission flux. The ability to obtain multiple secondary eclipses without data stacking means that we can quantify the climate variability of KELT-9b, for which only an upper limit was previously set on HD 189733b using Spitzer data \citep{agol10}.

One may use the secondary eclipse measurements to perform eclipse mapping \citep{dewit12,m12}.  The eclipse mapping technique is sensitive to the brightness temperature contrast on the planet dayside. Assuming a hotter region on the planet photosphere with a brightness temperature 1500 K larger than the continuum would yield a 30-ppm amplitude signal in both occultation ingress and egress in the CHEOPS bandpass.  We estimate that a photometric precision of 15 ppm over 2 minutes is required to achieve a 4-$\sigma$ detection of the ingress/egress features. We expect CHEOPS to achieve a precision of about 100 ppm over 2 minutes, which would require about 45 occultations to retrieve an optical eclipse map of that planet. These observations bear two caveats. First, the impact parameter of $b=0.17$ \citep{gaudi17} will negatively impact our ability to constrain the latitudinal location of the hot spot. Second, the level of stellar activity on short timescales ($\sim$minutes) probed at visible wavelengths will be critical to assess the feasibility of these observations. 

\section{Summary and future work}

\subsection{Summary}

In the current study, we have explored the atmospheric chemistry of ultra-hot Jupiters for which KELT-9b is an archetype.  In contrast to other studies \citep{arc18,kr18,loth18,man18,par18}, we do not assume chemical equilibrium in our calculations, but instead perform photochemical kinetics calculations, which explicitly quantify the effects of atmospheric mixing and photochemistry.  We then further explore the atmospheric chemistry using chemical-equilibrium calculations that include a set of metals.  Our findings include:
\begin{itemize}

\item Photochemistry, rather than atmospheric mixing, is the main driver for disequilibrium chemistry at low pressures.  Nevertheless, at pressures probed by optical and infrared spectroscopy at low resolution, chemical equilibrium is a reasonable approximation.

\item When analyzed in tandem, the abundances of carbon monoxide and water allows one to completely solve for C/H, O/H and temperature, provided that chemical equilibrium is assumed.

\item Metals in their atomic form are predicted to be observable.  In particular, atomic iron should be seen via a forest of optical and near-infrared lines using high-resolution ground-based spectrographs such as HARPS-N and CARMENES.

\end{itemize}

\subsection{Future work}

While we have argued that horizontal mixing does not significantly introduce chemical disequilibrium in ultra-hot Jupiters, a more rigorous investigation would involve running three-dimensional GCMs of KELT-9b-like exoplanets that include a treatment of disequilibrium chemistry (rather than assuming chemical equilibrium) and a varying specific heat capacity (due to the co-existence of atomic and molecular hydrogen in the same atmosphere).  Existing work in the published literature already hint that results from hot-Jupiter GCMs (executed at lower temperatures than for KELT-9b) may not carry over straightforwardly.  For example, \cite{showman13} studied the transition between moderate and intense stellar insolation using GCMs.  To lowest order, the zonal wind speeds are driven by the dayside-nightside temperature contrast, which is unknown for KELT-9b.  A higher dayside-nightside temperature contrast should lead to faster winds, but this cannot increase without bound \citep{fromang16}.  Under moderate stellar insolation, the usual flow pattern of a zonal jet appears (see \citealt{hs15} for a review).  Under intense stellar insolation, a combination of the damping of planetary-scale waves and frictional drag inhibit zonal jet formation and dayside-to-nightside flow.  It remains an open question if the atmospheric flow of ultra-hot-Jupiters occupy yet another regime and if these flows are capable of significantly driving the atmosphere away from chemical equilibrium.  Future work needs to elucidate the flow behavior and its influence on the chemistry of KELT-9b-like exoplanets in three dimensions.

\begin{acknowledgments}
We acknowledge financial support from the Swiss National Science Foundation, the European Research Council (via a Consolidator Grant to KH; grant number 771620), the PlanetS National Center of Competence in Research (NCCR), the Center for Space and Habitability (CSH) and the Swiss-based MERAC Foundation.
\end{acknowledgments}

\label{lastpage}


\begin{thebibliography}{99}

\bibitem[Agol et al.(2010)]{agol10} Agol, E., Cowan, N.B., Knutson, H.A., et al. \ 2010, ApJ, 721, 1861

\bibitem[Ag\'{u}ndez et al.(2014)]{agundez14} Ag\'{u}ndez, M., Parmentier, V., Venot, O., Hersant, F., \& Selsis, F. \ 2014, A\&A, 564, A73

\bibitem[Aldenius \& Johansson(2007)]{aj07} Aldenius, M., \& Johansson, S. \ 2007, A\&A, 467, 753

\bibitem[Arcangeli et al.(2018)]{arc18} Arcangeli, J., D\'{e}sert, J.-M., Line, M.R., et al. \ 2018, ApJL, 855, L30

\bibitem[Asplund et al.(2009)]{asp09} Asplund M., Grevesse N., Sauval A. J., Scott P. \ 2009, ARA\&A, 47, 481

\bibitem[Bell \& Cowan(2018)]{bc18} Bell, T.J., \& Cowan, N.B. \ 2018, ApJL, 857, L20

\bibitem[Broeg et al.(2013)]{broeg13} Broeg, C., Fortier, A., Ehrenreich, D., et al. \ 2013, Hot Planets and Cool Stars, Garching, Germany, Edited by Roberto Saglia, EPJ Web of Conferences, Volume 47, id.03005

\bibitem[Brogi et al.(2012)]{brogi12} Brogi, M., Snellen, I.A.G., de Kok, R.J., et al. \ 2012, Nature, 486, 502

\bibitem[Cooper \& Showman(2006)]{cs06} Cooper, C.S., \& Showman, A.P. \ 2006, ApJ, 649, 1048

\bibitem[Corr\'{e}g\'{e} \& Hibbert(2006)]{ch06} Corr\'{e}g\'{e}, G., \& Hibbert, A. \ 2006, ApJ, 636, 1166

\bibitem[de Wit et al.(2012)]{dewit12} de Wit, J., Gillon, M., Demory, B.-O., \& Seager, S. \ 2012, A\&A, 548, A128

\bibitem[Draine(2011)]{draine} Draine, B.T. \ 2011, Physics of the Interstellar and Intergalactic Medium (Princeton University Press)

\bibitem[Drummond et al.(2016)]{drummond16} Drummond, B., Tremblin, P., Baraffe, I., et al. \ 2016, A\&A, 594, A69

\bibitem[Drummond et al.(2018)]{drummond18} Drummond, B., Mayne, N.J., Manners, J., et al. \ 2018, ApJL, 855, L31

\bibitem[Ehrenreich et al.(2015)]{ehrenreich15} Ehrenreich, D., Bourrier, V., Wheatley, P.J., et al. \ 2015, Nature, 522, 459

\bibitem[Fortney(2005)]{fortney05} Fortney, J.J. \ 2005, MNRAS, 364, 649

\bibitem[Fromang et al.(2016)]{fromang16} Fromang, S., Leconte, J., \& Heng, K. \ 2016, A\&A, 591, A144

\bibitem[Gaudi et al.(2017)]{gaudi17} Gaudi, B.S., Stassun, K.G., Collins, K.A., et al. \ 2017, Nature, 546, 514

\bibitem[Grimm \& Heng(2015)]{grimm15} Grimm, S.L., \& Heng, K. \ 2015, ApJ, 808, 182

\bibitem[Guillot(2010)]{guillot10} Guillot, T. \ 2010, A\&A, 520, A27

\bibitem[Heng \& Demory(2013)]{hd13} Heng, K., \& Demory, B.-O. \ 2013, ApJ, 777, 100

\bibitem[Heng \& Showman(2015)]{hs15} Heng, K., \& Showman, A.P. \ 2015, AREPS, 43, 509

\bibitem[Heng(2016)]{heng16} Heng, K. \ 2016, ApJL, 826, L16

\bibitem[Heng \& Tsai(2016)]{ht16} Heng, K., \& Tsai, S.-M. \ 2016, ApJ, 829, 104

\bibitem[Heng(2017)]{heng17} Heng, K. \ 2017, Exoplanetary Atmospheres: Theoretical Concepts and Foundations (Princeton University Press)

\bibitem[Heng \& Kitzmann(2017)]{hk17} Heng, K., \& Kitzmann, D. \ 2017, MNRAS, 470, 2972

\bibitem[Husser et al.(2013)]{Husser} Husser, T.-O., Wende-von Berg, S., Dreizler, S., et al. \ 2013, A\&A, 553, A6
  
\bibitem[John(1988)]{john88} John, T.L. \ 1988, A\&A, 193, 189

\bibitem[Jones et al.(2013)]{jones13} Jones, A., Noll, S., Kausch, W., Szyszka, C., \& Kimeswenger, S. \ 2013, A\&A, 560, A91

\bibitem[Kaastra et al.(2008)]{k08} Kaastra, J.S., Paerels, F.B.S., Durret, F., Schindler, S., \& Richter, P. \ 2008, Space Science Reviews, 134, 155

\bibitem[Komacek \& Tan(2018)]{kt18} Komacek, T.D., \& Tan, X. \ 2018, AAS Research Notes (arXiv:1805.07415)

\bibitem[Kreidberg et al.(2018)]{kr18} Kreidberg, L., Line, M.R., Parmentier, V., et al. \ 2018, AJ, in press (arXiv:1805.00029v2)

\bibitem[Kuchner \& Seager(2005)]{ks05} Kuchner, M.J., \& Seager, S. \ 2005, arXiv:astro-ph/050421

\bibitem[Line et al.(2011)]{line11} Line, M.R., Vasisht, G., Chen, P., Angerhausen, D., \& Yung, Y.L. \ 2011, ApJ, 738, 32

\bibitem[Lothringer et al.(2018)]{loth18} Lothringer, J.D., Barman, T., \& Koskinen, T. \ 2018, arXiv:1805.00038v1

\bibitem[Madhusudhan(2012)]{madhu12} Madhusudhan, N. \ 2012, ApJ, 758, 36

\bibitem[Majeau et al.(2012)]{m12} Majeau, C., Agol, E., \& Cowan, N.B. \ 2012, ApJL, 747, L20

\bibitem[Malik et al.(2017)]{malik17} Malik, M., Grosheintz, L., Mendon\c{c}a, J.M., et al. \ 2017, AJ, 153, 56

\bibitem[Mansfield et al.(2018)]{man18} Mansfield, M., Bean, J.L., Line, M.R., et al. \ 2018, arXiv:1805.00424v1

\bibitem[Moses et al.(2011)]{moses11} Moses, J.I., Visscher, C., Fortney, J.J., et al. \ 2011, ApJ, 737, 15

\bibitem[Moses et al.(2013)]{moses13} Moses, J.I., Line, M.R., Visscher, C., et al. \ 2013, ApJ, 777, 34

\bibitem[Noll et al.(2012)]{noll12} Noll, S., Kausch, W., Barden, M., et al. \ 2012, A\&A, 543, A92

\bibitem[Nussbaumer \& Storey(1988)]{ns88} Nussbaumer, H., \& Storey, P.J. \ 1988, A\&A, 193, 327

\bibitem[Oreshenko et al.(2017)]{oreshenko17} Oreshenko, M., Lavie, B., Grimm, S.L., et al. \ 2017, ApJL, 847, L3

\bibitem[Parmentier et al.(2018)]{par18} Parmentier, V., Line, M.R., Bean, J.L., et al. \ 2018, arXiv:1805.00096v1

\bibitem[Peterson et al.(2017)]{peterson17} Peterson, R.C., Kurucz, R.L., \& Ayres, T.R. \ 2017, ApJS, 229, 23

\bibitem[Ryabchikova et al.(2015)]{ryabchikova15} Ryabchikova, T., Piskunov, N., Kurucz, R. L., et al. \ 2015, Physica Scripta, Volume 90, Issue 5, article id. 054005

\bibitem[Rimmer \& Helling(2016)]{rh16} Rimmer, P.B., \& Helling, Ch. \ 2016, ApJS, 224, 9

\bibitem[Seager(2010)]{seager10} Seager, S. \ 2010, Exoplanet Atmospheres: Physical Processes (Princeton University Press)

\bibitem[Schr\"{o}der \& Schmitt(2007)]{ss07} Schr\"{o}der, C., \& Schmitt, J.H.M.M. \ 2007, A\&A, 475, 677

\bibitem[Showman et al.(2013)]{showman13} Showman, A.P., Fortney, J.J., Lewis, N.K., \& Shabram, M. \ 2013, ApJ, 762, 24

\bibitem[Smith(1998)]{smith98} Smith, M.D. \ 1998, Icarus, 132, 176

\bibitem[Snellen et al.(2010)]{snellen10} Snellen, I.A.G., de Kok, R.J., de Mooij, E.J.W., \& Albrecht, S. \ 2010, Nature, 465, 1049

\bibitem[Solanki et al.(1992)]{s92} Solanki, S.K., Rueedi, I.K., \& Livingston, W. \ 1992, A\&A, 263, 312

\bibitem[Stevenson(2016)]{stevenson16} Stevenson, K.B. \ 2016, ApJL, 817, L16

\bibitem[Stock et al.(2018)]{stock18} Stock, J.W., Kitzmann, D., Patzer, A.B.C., \& Sedlmayr, E. \ 2018, MNRAS, in press (arXiv:1804.05010)

\bibitem[Tsai et al.(2017)]{tsai17} Tsai, S.-M., Lyons, J.R., Grosheintz, L., et al. \ 2017, ApJS, 228, 20

\bibitem[Tsai et al.(2018)]{tsai18} Tsai, S.-M., Kitzmann, D., Lyons, J.R., et al. \ 2018, ApJ, in press (arXiv:1711.08492)

\bibitem[Venot et al.(2013)]{venot13} Venot, O., Fray, N., B\'{e}nilan, Y., et al. \ 2013, A\&A, 551, A131

\bibitem[Venot et al.(2015)]{venot15} Venot, O., H\'{e}brard, E., Ag\'{u}ndez, M., Decin, L., \& Bounaceur, R. \ 2015, A\&A, 577, A33

\bibitem[Visscher(2012)]{vis12} Visscher, C. \ 2012, ApJ, 757, 5

\bibitem[Vidal-Madjar et al.(2004)]{vm04} Vidal-Madjar, A., D\'{e}sert, J.-M., Lecavelier des Etangs, A., et al. \ 2004, ApJL, 604, L69

\bibitem[Zahnle \& Marley(2014)]{zm14} Zahnle, K.J., \& Marley, M.S. \ 2014, ApJ, 797, 41

\end{thebibliography}
\end{document}